\documentclass[aps,prb,onecolumn,groupedaddress,preprint]{revtex4-1}
\usepackage{graphicx}  
\usepackage{dcolumn}   
\usepackage{float}
\usepackage{bm}        
\usepackage{hyperref}
\usepackage{amssymb}   
\usepackage{color}

\begin{document}

\title{Surface single-molecule dynamics controlled by entropy at low temperatures}

\author{J.~C.~Gehrig$^1$}
\author{M.~Penedo$^1$}
\author{M.~Parschau$^1$}
\author{J.~Schwenk$^1$}
\author{M.~A.~Marioni$^1$}
\author{E.~W.~Hudson$^{3,1}$}
\author{H.~J.~Hug$^{1,2}$}
\affiliation{$^1$Empa, Swiss Federal Laboratories for Materials Science and Technology, CH-8600 D\"{u}bendorf, Switzerland.}
\affiliation{$^2$Department of Physics, University of Basel, CH-4056 Basel, Switzerland.}
\affiliation{$^3$Department of Physics, Pennsylvania State University, University Park, PA 16802, USA.}
\email[Corresp. author. E-mail:~]{miguel.marioni@empa.ch}

\date{\today}


\maketitle

{
\bf
Configuration transitions of individual molecules \cite{Lauhon:1999fr,Baber:2008} and atoms \cite{Barth:2000} on surfaces are traditionally described with energy barriers and attempt rates using an Arrhenius law.
This approach yields consistent energy barrier values, but also attempt rates orders of magnitude below expected oscillation frequencies of particles in the meta-stable state \cite{Barth:2000,Heinrich:2002,Baber:2008,Marbach:2014}.
Moreover, even for identical systems, the measurements can yield values differing from each other by orders of magnitude \cite{Baber:2008,Jewell:2010}.
Using low temperature scanning tunnelling microscopy (STM) to measure an individual dibutyl-sulfide molecule (DBS) on Au(111), we show that we can avoid these apparent inconsistencies if we account for the relative position of tip apex and molecule with accuracy of a fraction of the molecule size.
Altering the tip position on that scale modifies the transition's barrier and  attempt rate in a highly correlated fashion, which on account of the relation between the latter and entropy results in a single-molecular enthalpy-entropy compensation \cite{Hanggi:1990,Marbach:2014}.
By appropriately positioning the tip apex the STM tip can be used to select the operating point on the compensation line and modify the transition rates.
The results highlight the need to consider entropy in transition rates of a single molecule, even at temperatures where entropy effects are usually neglected.
}

Observing transitions of molecules between metastable states provides insight into the dynamics underpinning many catalytic\cite{Bond:2000iba,Liu:2001hc} and biological processes\cite{Dunitz:1995,Villa:2000,Olsson:2011hw,Hanlumyuang:2014kl,Greenbaum:2014is,Dong:2015cn}.
We can restrict the phase space of the molecules by adsorbing them on a surface such that their transition kinetics becomes accessible to STM.
Thus for instance, it is possible to calculate hopping rates from series of images tracking the surface motion of Al atoms on Al(111) \cite{Barth:2000}, tetrapyridylporphyrin on Cu(111) \cite{Eichberger:2008}, tetraphenylporphirin monomers and dimers on Cu(111) \cite{Buchner:2011}, and cobalt(II) octaethylporphyrin on graphite \cite{Friesen:2012}.
Certain adsorbed molecules can also pivot around their bond to the substrate, and distinct rotational states can be imaged in sequence by STM, as shown for O$_2$ on Pt(111) \cite{Stipe:1998}, Cu phthalocyanine on C$_{60}$ \cite{Stohr:2001,Fendrich:2006} and zinc metalloporphyrin zinc tetra-(3,5-di-tert-butylphenyl)porphyrin on Ag(100) \cite{Vaughan:2006}.
In these types of transitions the rate of occurrence can also be obtained by counting individual transitions from telegraph noise in the current signal \cite{Stipe:1998,Tierney:2011,Schaffert:2013rsi}.
Such rate data is the basis for studying the temperature dependent physics of the transition, most commonly described in terms of an Arrhenius law \cite{Lauhon:1999fr} when the process is thermally activated.
In this context one can determine energy barriers to the transitions ($E^*$) and transition attempt rates ($A^*$) \cite{Lauhon:1999fr}.

The latter parameter is formally a constant, yet in various instances the measured values of $A^*$ depart from expected behavior in two ways:
First, $A^*$ measures consistently below molecular vibration frequencies in realistic adsorption surface potentials \cite{Chung:2006, Baber:2008};
Second, the values for the same system and experiment conditions can vary by orders of magnitude, notably in dibutyl-sulfide on Au(111) \cite{Baber:2008,Jewell:2010}.
In this work we look closer at the reasons for this puzzle, using an STM tip to modify and characterize transitions rates of a single-molecule model system.

We study the Arrhenius behavior of a dibutyl sulfide molecule adsorbed on Au(111) in the temperatures range 5 -- 10 K.
These molecules adsorb via their central sulfur atom, which becomes the pivot for molecular rotation, with the two butyl groups hopping between equivalent sites on the Au(111) surface.
\begin{figure*}[p]
\centering
\includegraphics[width=165mm]{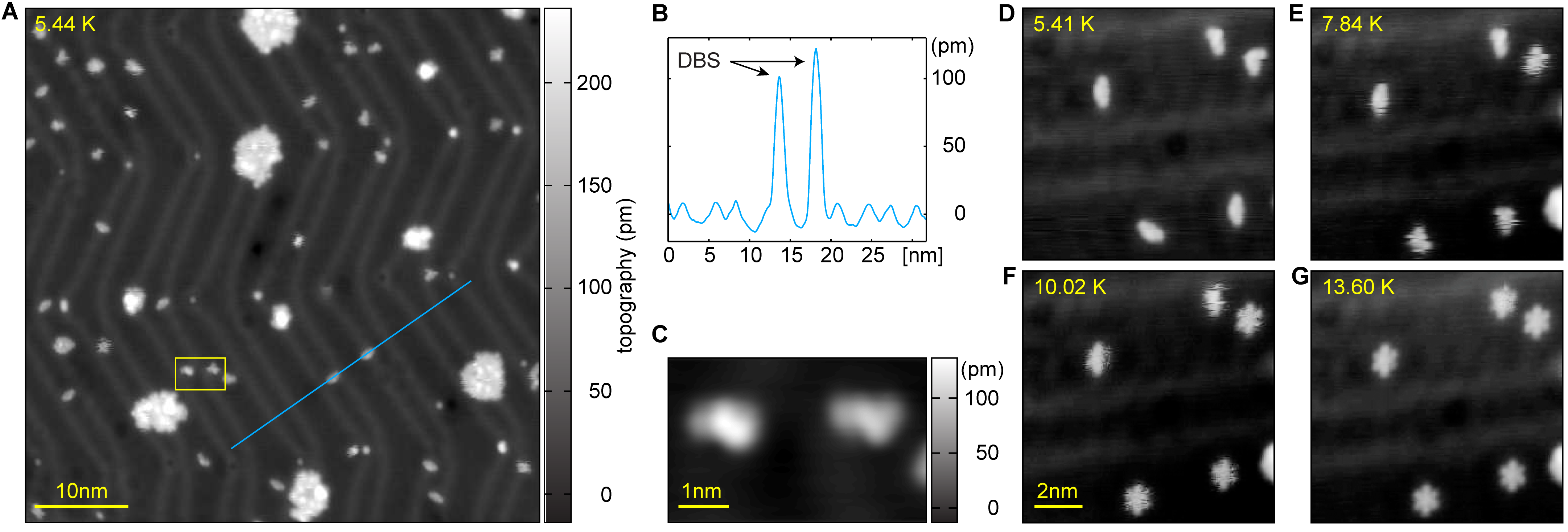}
\caption{
\textsf{\textbf{STM measurements of dibutyl sulfide molecules.}
\textbf{\textsf{A}} Overview image of dibutyl sulfide molecules on Au(111).
Isolated molecules decorate the surface.
\textbf{\textsf{B}} Section of \textbf{\textsf{A}} along the blue line, showing the height of the DBS molecule (distinct, large peaks) as well as the typical Au(111) herringbone structure (periodic, small peaks).
\textbf{\textsf{C}} Enlargement of the $5.2\times 3.3$\,nm$^2$ yellow-framed area of \textbf{\textsf{A}}.
\textbf{\textsf{D}}\,--\,\textbf{\textsf{G}} Series of topographs of isolated molecules taken at increasing temperatures.
At 5.41\,K molecules appear to be linear, whereas at 13.60\,K six lobes are visible consistent with the occupation of the different rotated configurations.
}}
\label{Fig.1}
\end{figure*}
We deposit dibutyl sulfide molecules at about 20\,K on a Au(111) substrate mounted on a cooled manipulator, and subsequently transfer the sample to our home-built, low-temperature UHV-STM/SFM system located inside a bath cryostat.
The temperature remains below 50\,K where molecular surface diffusion is suppressed sufficiently for many molecules to remain conveniently isolated (cf.~Fig.~1\textsf{A}).

For all STM imaging in this work we use a bias of 200\,mV and keep the (average) tunnelling current constant at 20\,pA, which amounts to a tunnelling resistance of 10\,G$\Omega$.
It is generally believed that the tip does not influence the dynamics of the molecule under such tunnelling conditions \cite{Baber:2008,Marbach:2014}.

STM measurements at 5.44\,K reveal isolated molecules appearing as linear objects of about 120\,pm height (Fig.~1\textsf{B}).
Their rotation by hopping can be induced with temperature (Fig.~1\textsf{D}-\textsf{G}) or inelastic tunnelling\cite{Baber:2008}.
Note that the latter mechanism would trigger molecule responses at a temperature independent rate.
In this work we use a bias of 200\,mV, which is well below the 375\,mV threshold for the inelastic tunnelling which was shown to excite a C-H stretching mode\cite{Jewell:2010} in DBS on Au(111).
Consistent with it the rotation hops of DBS rarely take place at temperatures below 6\,K, with a hopping rate sufficiently small for the butyl groups to remain in the same position during the recording time of an STM image, as shown in Figs.~1\textsf{C} and \textsf{D}.
As the hops become more frequent with temperature increasing from 5.41\,K to 13.60\,K (Figs.~1\textsf{D}-\textsf{G}), the butyl groups occupy more than one position during the image scan, distorting the linear appearance of the molecules.
At 13.60\,K the butyl groups hop so quickly that the molecule appears to have a hexagonal shape (Fig.\,1\textsf{G}).

In contrast to prior work \cite{Baber:2008,Jewell:2010}, high resolution data ($\approx 50\,$pm square per pixel) are taken over a $5.2$\,nm$\times 3.3$\,nm area at six different temperatures ($T_{i=1,...,6} = 7.39$, 7.87, 8.38, 8.9, 9.48 and 10.05\,K) using a slow z-feedback (feedback parameters: $P=10^{-5}$\,nm/nA, $I=1$\,ms) at a scan rate of 400\,s per line (2\,s per pixel in each of the forward/backward direction).
This leads to $100\times 64$\,point topographs (e.g. Fig.~1\textsf{C}) coupled with 3000\,point tunnel current traces (e.g.~Fig.~2\textsf{A}) sampled at 1.5\,kHz, beyond the bandwidth of our current preamplifier (1.1\,kHz for a current-to-voltage conversion gain of $10^9$\,V/A).
A posteriori analysis of this latter data allows us to obtain the number of hops at each pixel $N(x,y,T)$; the hopping rate is simply this divided by the time spent on each pixel, i.e.~$k(x,y,T) \equiv N(x,y,T)/2\,\rm{s}$.
The scan range of the piezo scanner increases with temperature and can lead to small changes in the scale of the images, a fact that we account for by separate calibration measurements carried out over the relevant temperature range (not shown).
It is also necessary to correct the drift between data sets acquired at different temperatures. This we do by laterally shifting the images relative to the 8.9\,K image so as to maximize the cross correlation between the current error ($I(t)-20$\,pA, not shown) images.
Alignment with the topography channel leads to nearly identical results.

In figures~2\textsf{B}-\textsf{G} we show representative hopping-rate measurements made on the left molecule of Fig.~1\textsf{C}.
The figure highlights both a general trend of hopping rate increasing with temperature and the rate's strong (one order of magnitude) dependence on the position of the STM tip relative to the molecule.
The patterns are not random, and lack the symmetry of the highest-temperature images (e.g.~13.6\,K image in Figure 1G, acquired on a different dibutyl sulfide molecule with a different tip).
\begin{figure}[p]
\centering
\includegraphics{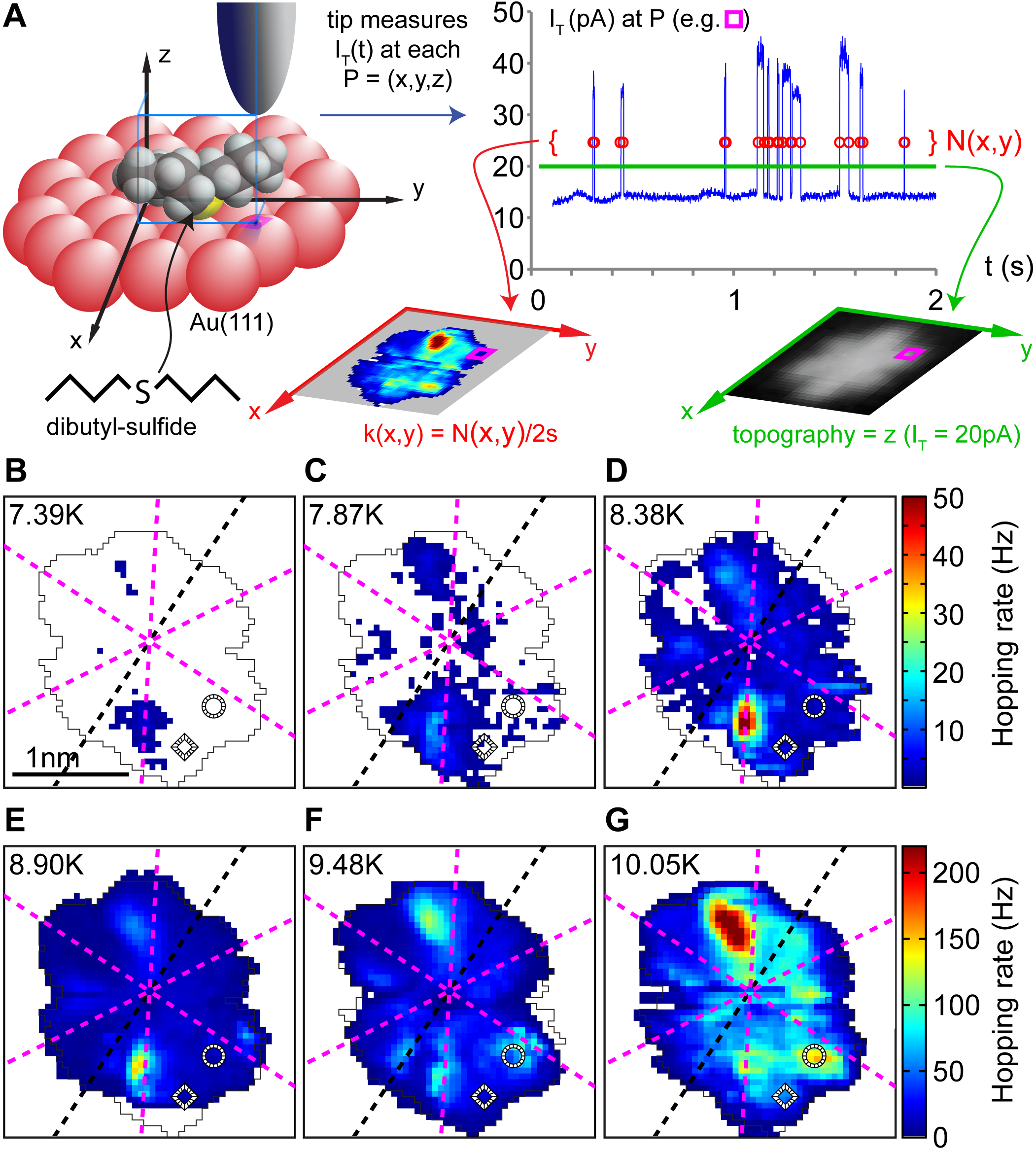}
\caption{
\textsf{
\textbf{Hopping rate measurement and analysis.}
\textbf{\textsf{A}} Measurement principle: A DBS molecule adsorbed on the Au(111) surface (top left) is scanned at a rate of 400\,s/line (both directions).
$z$-feedback keeps the average tunnelling current constant at 20\,pA and results in the topography image (bottom right).
Telegraph noise recorded at a high measurement bandwidth (top right) allows identifying and counting the $N$ molecular hops occurring over the 2\,s span that the tip spends over each position $(x,y)$, and results in the hopping rate $k(x,y,T)$ (bottom left).
\textbf{\textsf{B}}--\textbf{\textsf{G}} Maps of the hopping rate $k(x,y,T)$ highlight the influence of both temperature and relative molecule-tip position $(x,y)$.
Note that in these and all subsequent spatial maps, the color white indicates regions where no hopping was discerned using this technique.
The black/white striped ring and diamond markers identify two specific positions $r_1$ and $r_2$ for further analysis in Fig.~3.
The dashed black line is parallel to the nearby herring bone reconstruction line, while the dashed magenta lines are the perpendicular and three-fold rotations.
The black contour outlines the region utilized for further analysis, containing non-zero rates for at least 3 temperatures.
The hexagonal shape of this contour is reminiscent of the typical hexagonal shaped appearance of STM topography map of the molecule typically occurring at higher temperatures (Fig. 1G).
}
}
\label{Fig.2}
\end{figure}

Given the aforementioned measurement conditions\cite{Baber:2008} it is surprising that the transition rate should depend on tip-position.
Yet the fact that it does indicates that the tip influences the system, although the effect becomes evident only when looking at the transition rates, and may have been overlooked until now.
Consistent with this it is reasonable to attribute the symmetry of the rate patterns at least in part to the (unknown) arrangement of the atoms of the tip apex, which will generally not have the symmetry of the substrate at the adsorption point.

The consequence of these considerations is that, a priori, we must treat the tip as an integral part of the system, inasmuch as it can modify the thermodynamic potentials.
Possibly, however, these tip-induced modifications could be used to extract information of the system dynamics.
To assess the extent of the tip-influence and establish the implications for the dynamics of molecules on surfaces in general, we analyze how the hopping rate $k(x,y,T)$ changes with temperature $T$ at each $(x,y)$ tip-position.
After ensuring the rate maps measured at each temperature are in registry (see Supporting material) we can carry out the comparison as indicated schematically in Fig.~3\textsf{A}.
Examples from selected positions in Fig.~3\textsf{B} indicate that the behavior is phenomenologically described by an Arrhenius equation,
\begin{equation}
k(x,y,T) = A^*(x,y)\cdot \exp \left(-\frac{E^*(x,y)}{k_BT}\right),
\label{eq:Arrhenius}
\end{equation}
where $k_B$ is Boltzmann's constant, $E^*(x,y)$ is an apparent energy barrier height, and $A^*(x,y)$ is an apparent attempt rate.
$E^*(x,y)$ and $A^*(x,y)$ describe the lines fitting the data in Fig.~3\textsf{B} and are given inside that figure's panel.

For the complete set of scanned $(x,y)$ positions the $E^*(x,y)$ and $A^*(x,y)$ fitting the measured $k(x,y,T)$ with equation~(\ref{eq:Arrhenius}) result in maps depicted in Fig.~3\textsf{C} and \textsf{D} respectively.
Panels \textsf{E} and \textsf{F} indicate the corresponding errors.
(Panel \textsf{G} indicates the number of points used for the fit.)
For most tip positions the Arrhenius law is a good fit to the data, meaning that the transitions are essentially thermally activated.
\begin{figure}[p]
\centering
\includegraphics{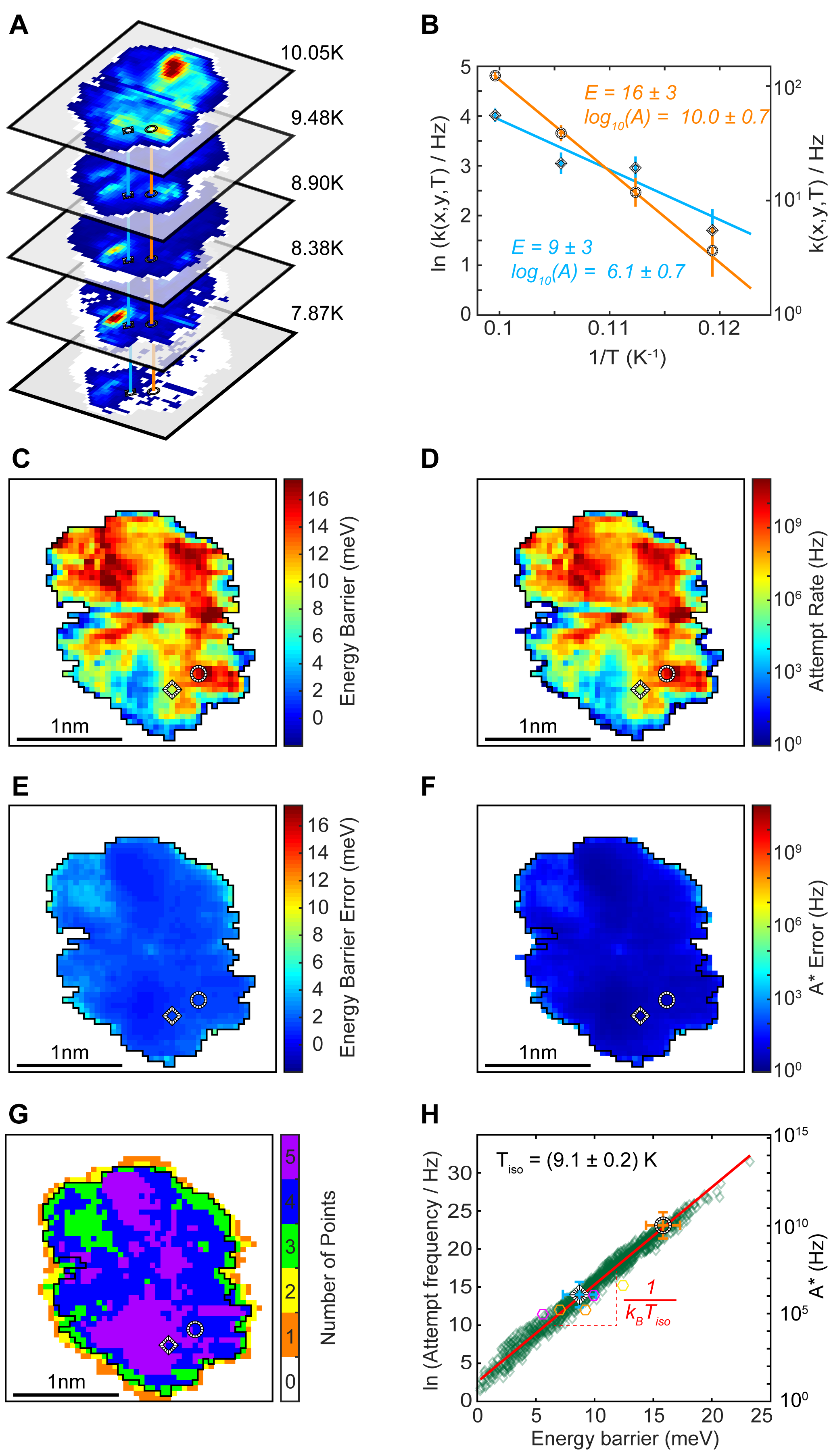}
\caption{
\textsf{\textbf{Arrhenius law parameters from fits to the hopping rate $k(x,y,T)$ of Fig.~2B--G}.
\textbf{\textsf{A}} Schematic of the procedure followed to obtain sets of position dependent hopping rates $k(x,y,T)$ for fitting with equation~(\ref{eq:Arrhenius}).
The black/white striped ring and diamond markers identify two selected $(x,y)$ positions across temperatures.
\textbf{\textsf{B}} Arrhenius plots of $k(x,y,T)$ for the two positions corresponding to the striped markers in \textbf{\textsf{A}} and Fig.~2.
\textbf{\textsf{C}} and \textbf{\textsf{D}} Apparent energy barrier $E^*(x,y)$ and apparent attempt rate $A^*(x,y)$ from fitting equation~(\ref{eq:Arrhenius}) to each $(x,y)$ hopping rate temperature series.
\textbf{\textsf{E}} and \textbf{\textsf{F}} Error of the measurement of \textbf{\textsf{C}} and \textbf{\textsf{D}}, respectively.
\textbf{\textsf{G}} Number of points available in the temperature series of position-dependent hopping rates for each $(x,y)$.
\textbf{\textsf{H}} Compensation effect between $E^*$ and $A^*$ from \textbf{\textsf{C}} and \textbf{\textsf{D}}, from which an isokinetic temperature $T_{\rm{iso}}$ follows.
Values from literature are indicated by hexagons in yellow\cite{Baber:2008} (fcc), magenta\cite{Jewell:2010} (fcc) and orange\cite{Jewell:2010} (hcp), respectively.
}}
\label{Fig.3}
\end{figure}

Several observations are noteworthy.
First, the effective energy barrier height $E^*(x,y)$ spans an unexpectedly large range from a few meV to about 20\,meV (Fig.~3\textsf{F}).
Second, as we found with the rate, also the energy barrier depends on tip-position, showing that the overall energy landscape probed by the adsorbed molecule is strongly influenced by the presence of the tip.
And third, the attempt rate $A^*(x,y)$ varies over 10 orders of magnitude over the scanned relative positions, but its map (Fig.~3\textsf{D}) exhibits a close resemblance to that of $E^*(x,y)$ (Fig.~3\textsf{C}), suggesting a fundamental underlying connection between the two.

Taking all pairs of $\ln(A^*(x,y))$ and $E^*(x,y)$ and plotting them in Fig.~3\textsf{H} reveals a clustering of the data on a straight line.
Note that even this subset spans an attempt-rate range of more than five orders of magnitude.
Hexagons indicate the location on the plot of data from literature for rotation of dibutyl sulfide on Au(111), and the striped markers correspond to the lines presented in Figure 3\textsf{B}.

Taking $A^*$ in Hz we can fit the data in Fig.~3\textsf{F} with the line
\begin{equation}
\ln A^*(x,y) = \frac{E^*(x,y)}{E_{MN}} +\ln A_{00},
\label{eq:comp-law-EMN}
\end{equation}
where the parameters are the energy $E_{MN} = 0.78 \pm$0.02$\,\rm{meV}$ and $\ln A_{00} = 2.5 \pm 0.3$.

Equation (\ref{eq:comp-law-EMN}) expresses for the single dibutyl-sulfide molecule what is known e.g.~for families of related chemical reactions \cite{Boisvert:1995} as the Cremer-Constable- or Meyer-Neldel rule (see Refs.~\citenum{Bond:2000iba,Liu:2001hc} for a review).
We can see that the role of the STM tip is to modify the conditions in which the transitions take place.
Specifically, for each relative tip-molecule position the DBS rotation transition encounters a different energy barrier height (which is connected to the attempt rate).
But the transition itself is not barrier-less and is thermally activated.
Replacing $A^*(x,y)$ in Equation 1 by the expression in Equation 2 reveals that at $T = T_{iso} = E_{MN}/k_B = 9.1 \pm 0.2\,$K the rates become position-independent, i.e.~$k(x,y) = \ln A_{00}$.
It has been pointed out by Boisvert et al.\cite{Boisvert:1995} that the Meyer-Neldel energy is expected to be of the order of the energy excitations available in the energy reservoir.

The variability of $A^*(x,y)$ over orders of magnitude becomes explicitly apparent by writing equation (\ref{eq:Arrhenius}) in terms the canonical multidimensional transition state theory result \cite{Hanggi:1990},
\begin{equation}
k(x,y,T) = \left[\frac{k_B T}{h}\frac{Z^{\ddagger}(x,y)}{Z_0(x,y)}\right] \exp \left(-\frac{\Delta H^\ddagger(x,y)}{k_BT}\right)
\label{eq:3.17}
\end{equation}
in which the tip-position dependent transition state (metastable state) partition function $Z^{\ddagger}(x,y)$ ($Z_0(x,y)$) and enthalpy $\Delta H^{\ddagger}(x,y)$ are included.
Equation~(\ref{eq:3.17}) can be shown to be equivalent to a theory of the escape rate of a particle trapped in a one-dimensional potential and coupled bilinearly to a bath of an infinite set of harmonic oscillators \cite{Hanggi:1990}.
More specifically we can define an entropy difference between the metastable and transitions as $\Delta S^{\ddagger}(x,y) = k_B\ln (Z^{\ddagger}(x,y)/Z_0(x,y))$, of which the apparent attempt rate $A^*(x,y)$ depends exponentially:
\begin{equation}
A^*(x,y) \rightarrow A(x,y,T) :=  \frac{k_B T}{h}\exp \left(\frac{\Delta S^{\ddagger}(x,y) }{k_B}\right).
\label{eq:A-rate}
\end{equation}
Thus, using the logarithmic form of the Eyring equation, which accounts for the weak temperature dependence of $A(x,y,T)$, we can obtain the enthalpy and entropy maps of Fig.\,4\textsf{A} and \textsf{B}.
Note that $\Delta S^{\ddagger}(x,y)$ varies between $-2.2$ and $0$\,meV/K and has the same pattern as $\Delta H^{\ddagger}(x,y)$ which expresses an enthalpy-entropy compensation.
Yet a better point of comparison is the thermodynamic potential obtained by multiplying $\Delta S^{\ddagger}(x,y)$ with a relevant temperature.
For this case we chose the isokinetic temperature $T_{iso}:=E_{MN}/k_B = 9.1\pm 0.2\,$K.
This amounts to a rescaling of the map in Fig.\,4\textsf{B}, as indicated by the right scale bar.
\begin{figure}[p]
\centering
\includegraphics{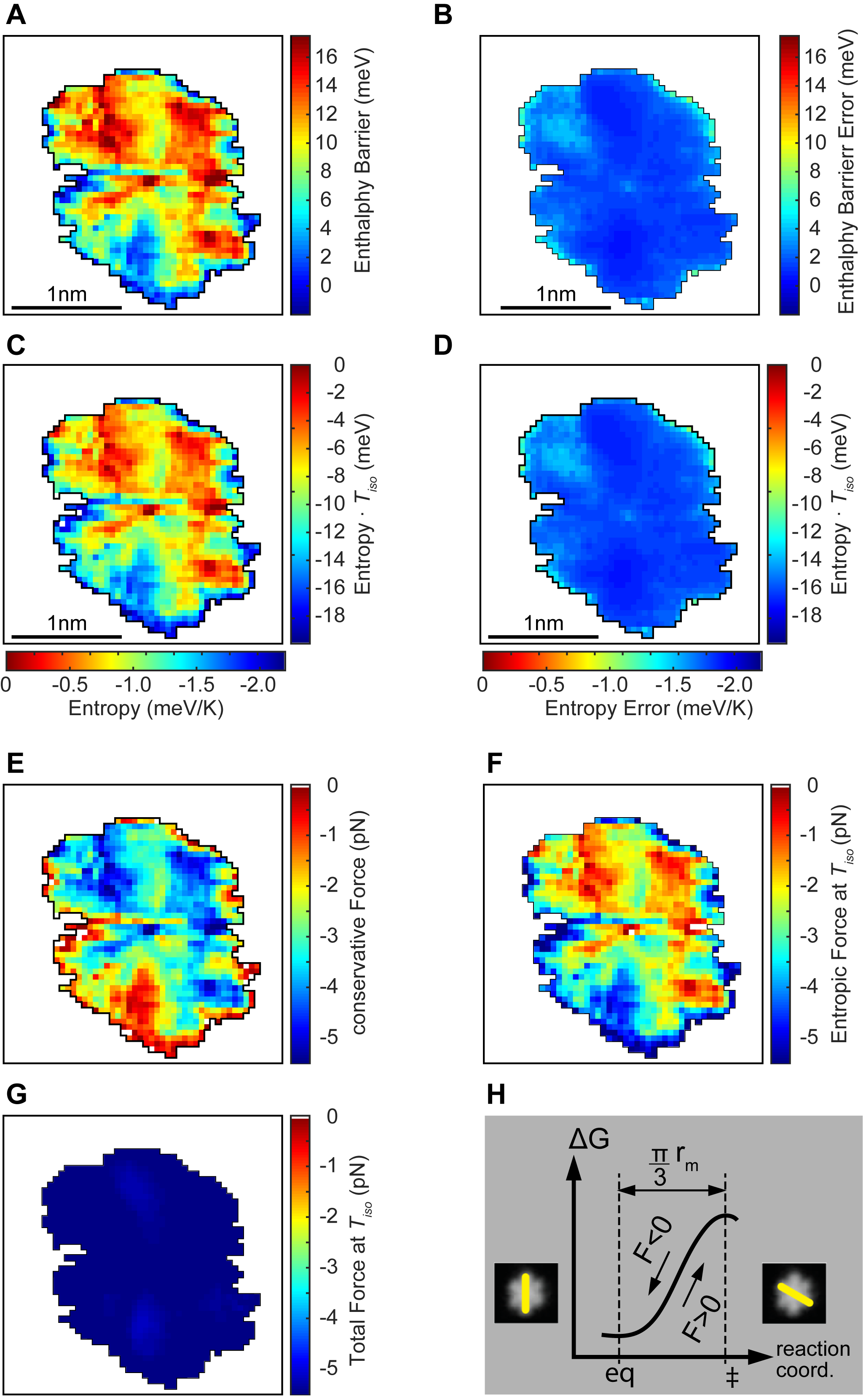}
\caption{
\textsf{
\textbf{Enthalpy and entropy of the DBS molecule derived from the data in Fig.~2B-G and equations~(\ref{eq:Arrhenius}) and (\ref{eq:A-rate}), and the resulting forces.}
\textbf{\textsf{A}} Map of the enthalpy change $\Delta H^{\ddagger}$ obtained by fitting the rate maps from Fig.\,3\textsf{A} with the logarithmic form of equation~(\ref{eq:Arrhenius}) using equation~(\ref{eq:A-rate}) for the (weakly) temperature dependent attempt rate, and $\Delta H^{\ddagger}(x,y)\equiv E^*(x,y)$.
\textbf{\textsf{B}} Corresponding error.
\textbf{\textsf{C}} Entropy difference $\Delta S^{\ddagger}$ from the fit of {\bf \textsf{A}}.
The right hand scale bar is a scale of the corresponding thermodynamic potential at $T_{\rm iso}$, which can be compared with {\bf \textsf{A}}.
\textbf{\textsf{D}} Corresponding error.
\textbf{\textsf{E}} Conservative force, derived from {\bf \textsf{A}}.
\textbf{\textsf{F}} Entropic force at $T_{\rm iso}$, derived from {\bf \textsf{C}}.
\textbf{\textsf{G}} Sum of conservative and entropic forces at $T_{\rm iso}$, from {\bf \textsf{E}} and {\bf \textsf{F}}.
\textbf{\textsf{H}} Schematic of the free energy change along the transition axis, indicating the sign convention for the forces $F$ (conservative or entropic) on the molecule.
}}
\label{Fig.4}
\end{figure}

It becomes clear that apart from the similarity of the patterns in Figs.\,4\textsf{A} and \textsf{B}, the data $T_{{\rm iso}}\Delta S^{\ddagger}$ is in the same range as $\Delta H^{\ddagger}$ (see right scale bar of Fig.\,4\textsf{B}).
Enthalpy and entropy have comparable effects on the dynamics of the DBS molecule rotation on the Au(111) surface.

Marbach et al. \cite{Marbach:2014} suggested that entropy can be important in such a process, but also pointed out that its contribution should become negligible at low temperatures.
The latter statement does not contradict our finding, but calls for a more careful assessment of the sources of entropy.
Experimental evidence suggests that configurational entropy, increasing with the number of degrees of freedom, could be behind the close proximity of the temperatures at which dialkyl-sufides with different chain lengths rotate \cite{Tierney:2011}.
Accordingly, if the STM tip constrained the possible locations of the butyl chains of the DBS molecule (analogously to the lateral confinement of metal-organic meshes in the work of Palma et al.\cite{Palma:2015}), then a concurrent change in entropy would follow.
However, it would not be a priori follow that the entropy change compensate the enthalpy (Fig.~3\textsf{F} and maps 4\textsf{B},\textsf{A}).

That this compensation nevertheless takes place in many instances was discussed by Yelon et al. \cite{Yelon:1990}, who placed its origin in multi-excitation entropy.
Their model applies to cases where energy barrier is much larger than the typical thermal fluctuation energies available, a condition which is given for our experiment.
In essence, a transition necessitates a potentially large number of small-energy excitations (of order $k_BT \approx 0.5$\,meV) to co-occur, so that the barrier is overcome.
Yelon et al.~then argue that {\em a large number of excitations may be collected in a large number of ways} and that this multiplicity gives rise to entropy\cite{Yelon:2006}.

From this perspective, any change in the barrier height induced by the tip would be followed by a compensation of the multi-excitation entropy.
It would be observed when other contributions to entropy do not dominate, and thus especially at low temperatures.

Lastly, comparing the conservative forces acting on the molecule with the emergent entropic forces shows the latter must be included in correct descriptions of the transition dynamics, even at low temperatures.
To see that, we can take the apparent barrier height $\Delta H^{\ddagger}$ and the potential $T_{{\rm iso}}\Delta S^{\ddagger}$, and divide them by an effective distance $r_M\times\pi/3\simeq 0.5$\,nm, where $r_M = 0.5\,$nm approximates the radius of the molecule.
The resulting conservative and emergent (entropic) force maps, shown in Fig.~4\textsf{E}--\textsf{F} span a comparable range of $-5.5$ -- 0\,pN.
These values are all about 100 times smaller than those required to push atoms across surfaces \cite{Stroscio:2006,Ternes:2008ka}, which can be understood by the relative weakness of the van der Waals interaction between the butyl group and the Au(111) surface compared to metallic bonds.
Because the values are much smaller, the transitions are thermally activated.

At $T = T_{\rm iso}$ the sum of the force maps in Fig.~4E-F is approximately independent of tip position (Fig.~4G), which illustrates the compensation of enthalpy and entropy.

Our work shows that entropic contributions to the dynamics of atoms and molecules during surface transitions and manipulation cannot be disregarded, even at low temperatures, and gives an example of how STM could be used to measure and modify the entropy of atomic or molecular-scale systems and their interactions with tip and substrate.
The STM tip can be used at a level of interaction with the molecule where the molecular transitions are still thermally activated, but the barriers are modified by the tip.
Thus different regimes of the thermally activated dynamics of the transitions can be probed by appropriately selecting the tip position, but the compensation itself is independent of the means by which the barrier height is modified.
It is therefore important that future models to describe the transition dynamics of surface-adsorbed molecules at low temperatures account for the transfer between substrate, molecule {\em and} tip of energies a fraction of the size of the transition barrier height.

{\noindent \bf \textsf{References}}

{\noindent \bf \textsf{Acknowledgments}}

{
\small
Support from the Swiss National Science Foundation, and Empa is hereby gratefully acknowledged.
We thank S.M. Vranjkovic for strong support in designing and constructing the STM utilized in this study, and acknowledge financial support from the EU FP7 program (MDSPM project).
We thank K.-H. Ernst for supplying the molecules and various scientific discussions on molecular motion on surfaces.
}

{\noindent \bf \textsf{Author contributions}}

{\small
J.C.Gehrig and M.Penedo carried out the measurements, and set up the STM together with M.Parschau, E.W.Hudson, J.Schwenk and H.J.Hug.
J.C.Gehrig, M.Penedo, M.A.Marioni, H.J.Hug, and E.W.Hudson analyzed the results.
M.A.Marioni, H.J.Hug and E.W.Hudson wrote the manuscript.
All authors discussed and commented the manuscript.
}

\end{document}